\documentclass[conference]{IEEEtran}
\IEEEoverridecommandlockouts
\usepackage{cite}
\usepackage{amsmath,amssymb,amsfonts}
\usepackage{algorithmic}
\usepackage{graphicx}
\usepackage{textcomp}
\usepackage{xcolor}
\usepackage{mhchem}
\usepackage{siunitx} 
\usepackage{balance}
\def\BibTeX{{\rm B\kern-.05em{\sc i\kern-.025em b}\kern-.08em
    T\kern-.1667em\lower.7ex\hbox{E}\kern-.125emX}}
\begin{document}

\title{Diffusive Molecular Communication with a Spheroidal Receiver for Organ-on-Chip Systems \\
%{\footnotesize \textsuperscript{*}Note: Sub-titles are not captured in Xplore andshould not be used}
\thanks{\copyright 2023 IEEE.  Personal use of this material is permitted.  Permission from IEEE must be obtained for all other uses, in any current or future media, including reprinting/republishing this material for advertising or promotional purposes, creating new collective works, for resale or redistribution to servers or lists, or reuse of any copyrighted component of this work in other works.}
}

\author{\IEEEauthorblockN{Hamidreza Arjmandi$^1$, Mohamad Zoofaghari$^2$, Mitra Rezaei$^1$, \\
		Kajsa Kanebratt$^3$, Liisa Vilen$^3$, David Janzen$^{4}$, Peter Gennemark$^{3,5}$, and Adam Noel$^1$ }
\IEEEauthorblockA{\textit{$^1$School of Engineering, University of Warwick, Coventry, UK.} \\
\textit{$^2$Electrical Engineering Department, Yazd University, Yazd, Iran.}\\
\textit{$^{3}$Drug  Metabolism  and  Pharmacokinetics,  Research  and  Early  Development, Cardiovascular,  
	Renal  and  Metabolism  (CVRM),  }\\
	 \textit{Bio-Pharmaceuticals, R\&D, AstraZeneca, Gothenburg, Sweden.}\\
	 	 \textit{$^{4}$Clinical Pharmacology and Quantitative Pharmacology, Clinical Pharmacology and Safety Sciences,}\\
	  \textit{AstraZeneca AB R\&D Gothenburg, Gothenburg, Sweden.}\\
	 \textit{$^{5}$Department of Biomedical Engineering, Link\"oping University, Sweden.}}
}

\maketitle

\begin{abstract}
Realistic models of the components and processes are required for molecular communication (MC) systems. In this paper, a spheroidal receiver structure is proposed for MC that is inspired by the 3D cell cultures known as \textit{spheroids} being widely used in organ-on-chip systems. A simple diffusive MC system is considered where the spheroidal receiver and a point source transmitter are in an unbounded fluid environment. The spheroidal receiver is modeled as a porous medium for diffusive signaling molecules, then its boundary conditions and effective diffusion coefficient are characterized. It is revealed that the spheroid amplifies the diffusion signal, but also disperses the signal which reduces the information communication rate. Furthermore, we analytically formulate and derive the concentration Green's function inside and outside the spheroid in terms of infinite series-forms that are confirmed by a particle-based simulator (PBS). 
\end{abstract}
\section{Introduction}
Molecular communication (MC) is a bio-inspired mechanism that is envisioned to realize micro- and nano-scale communication systems using molecules as information carriers \cite{b0}. 
%%%%add a sentence about why we should study molecular communication's part regarding their applications
%There are several models of MC receivers in the literature. The receivers in most of these models are viewed as ideal and simple, which does not reflect how cells deal with molecules in nature.
Despite many efforts by the MC community to model the components of MC systems, more realistic models are required. Elements and structures of \textit{in vitro} environments such as organs-on-chip could be used to improve model realism, potentially contribute to MC research and development, and to provide mechanistic insight into the biology of the organs (on the chip).

% necessary to to model and understand the physiology of organs, inter-organs communication, and the drug effectiveness and evaluation.

%In light of the importance of studying MC receivers, this research aims to provide a practical MC receiver model appropriate to use in the mentioned application.
Existing literature over-simplifies the MC receiver in a biological environment and does not sufficiently account for the interactions of signaling molecules with the environment and the biological or biosynthetic receivers and transmitters (i.e., cells). 
The passive receiver, in which information molecules freely diffuse in the receiver's space and the movement of molecules is not affected, is the most common model used in previous works \cite{b5}. Such a simple model is often used to facilitate analysis of other aspects of a MC system, e.g., the environment boundary \cite{Arjmandi19}. 
%This model can be used for hydrophobic and small molecules, such as dissolved gases and steroid hormones, which have the ability to cross the cell membrane readily \cite{b6}, \cite{b7}.
The passive model may be relevant for small and hydrophobic molecules (which are repelled from water molecules) that easily pass through a cell membrane. 
%After passing through the membrane, these molecules either directly activate intracellular enzymes or bind to intracellular receptor proteins. 
However, most extracellular molecules are too large or too hydrophilic to traverse the cell membrane \cite{b6} and also cells usually react with signaling molecules (either directly on the surface or in the intracellular environment). 

To address the limitations of the passive receiver model, some works have considered a reception mechanism across the receiver (cell) membrane activating a signaling pathway (i.e., a series of chemical reactions controlling a cell function). These works have studied the effects of various types of reactions across the membrane, as well as the size, number, spatial distribution, and/or saturation of the receptors on the cell membrane \cite{b8, b9,b11,b13,b22,b23,b15}.

Previous works have considered the receiver as a single cell (or machine) whose surface is reacting with molecules. However, cells do not normally live in isolation but in populations with other cells of the same or of different types. This is true \textit{in vivo} but also in many \textit{in vitro} systems. In particular, tissues and tumors in multi-cellular organisms and biofilms of microorganisms are common natural instances whereas spheroids, organoids, tumoroids, and cell islets are well-known instances in biological experimental setups. This inspires the design of MC transceivers based on a population of (biological or biosynthetic) cells. In \cite{Aziz19}, the authors consider an implantable drug delivery system inside a spherical tumor microenvironment releasing drug molecules and obtain the spatiotemporal drug concentration profile in the surrounding tissue.

One realistic receiver for MC is a spheroid structure, which is a 3D cell aggregation in a spherical shape that is widely used in organ-on-chip systems. Spheroids are constructed by various methods aiming to emulate the internal physiological activities of an organ \cite{b16}. %This compact spheroidal structure forms when the spontaneous aggregation of cells occurs, and the integrins on their surface bind to the extracellular matrix of their nearby cells \cite{b18}. 
In this paper, we consider a diffusive MC system with a spheroidal receiver and a point source transmitter. We model the spheroidal receiver as a porous medium for the diffusion signal and characterize its effective diffusion coefficient and implied boundary condition. We introduce and characterize two important processing features of the spheroid: \textit{amplification} and \textit{dispersion} of the diffusion signal. Furthermore, we obtain the concentration Green's function in an unbounded environment in the presence of the spheroidal receiver. Our results are confirmed by a particle based simulator (PBS).

%For instance, in targeted drug delivery systems, the drug-carrying transmitter must release the drug molecules toward the targeted cancerous cells, which are usually placed in a cluster and create a porous structure called a spheroid.  Furthermore, in the organ-on-a-chip, which simulates the internal physiological activities of an organ by culturing its cells on a microfluidic device, artificially spheroids constructed from different methods described in \cite{b16} are used as receivers to screen anti-cancer drugs.
%%examine the effects of drug treatments on them. %drug screening spheroids can be utilized for anti-cancer drug screening.
%The authors in \cite{b17} considered a tumor an aggregation of diseased cells, modeling it as one effective spherical absorbing receiver without considering its porous structure. 
%
%In this paper, with the aim of proposing a practical model, we analyze an MC system with a point source transmitter and a spheroidal receiver composed of several cells arranged close to each other with void spaces between them. We also obtain the close-form expression for the CIR of the system and verify it by particle-based simulation (PBS). We find that ...

The rest of the paper is organized as follows. Section II describes the spheroid structure and diffusive MC system. In Section III, the spheroid Green's function is provided. Results and discussions are presented in Section IV. Finally, Section V concludes the paper.

\section{System Model}
\subsection{Spheroid model}
 A spheroid with radius $R_s$ formed by $N_c$ cells is considered. The spheroid interior space is comprised of the cells and the void space between the cells (i.e., extracellular environment). Given that the volume of a cell within the spheroid is $V_c$, the volumes of the cell matrix and the void space inside the spheroid are given by $V_cN_c$ and $V_s-V_cN_c$, respectively, where $V_s=\frac{4}{3}\pi R_s^3$ is the spheroid volume. 

We model the spheroid structure as a porous medium with volume $V_s$ whose porosity, $\epsilon$, is defined as the ratio of the void space to the whole spheroid volume, i.e., 
\begin{align}\label{eps}
\epsilon=1-\frac{N_cV_c}{V_s}.
\end{align}
%The avergae distance between two neighber cells is given by

We assume that the spheroid is in a fluid medium that surrounds it and fills its void space. The diffusive signaling molecules of type $A$ in the medium can diffuse into the void space of the spheroid and stimulate the spheroid's cells through a transmembrane mechanism, e.g., by binding to receptors. %We define the sensing region of a cell as the space around the cell where the cell's receptors are impacted by the signaling molecules $A$. The size of the sensing region is of the same order of magnitude of the cell volume denoted by $V_c$ in liters ($l$).\footnote{To justify this assumption, one may assume a transparent cell that counts molecules fallen in its volume.} 

%For the sake of generality, we do not discuss the source of $A$ molecules in the environment or its concentration profile. 
Inside the spheroid, the molecules diffuse via the curved paths of the void space among the cells, which leads to a shorter net displacement of the molecules in a given time interval. Thus, macroscopic diffusion within the spheroid effectively differs from the diffusion within the free fluid outside the spheroid. Since the molecules traverse a shorter net path within the spheroid, the effective diffusion coefficient is smaller than the diffusion coefficient in the free fluid medium and molecules are more likely to be observed and sensed by the spheroid's cells. Given the diffusion coefficient $D$ for molecules $A$ in the free fluid, the effective diffusion coefficient within the spheroid is given by
\begin{align}
D_{\rm eff}=\frac{\epsilon}{\tau}D,
\end{align}
where $\tau$ is the tortuosity, a measure of the transport properties of the porous medium, and is modeled as a function of spheroid porosity as $\tau=\frac{1}{\epsilon^{0.5}}$ \cite{Sharifi19}.

To describe the environment geometry, we use the spherical coordinate system where $(r,\theta,\varphi)$ denote radial, elevation, and azimuth coordinates, respectively. At the border of the two diffusive environments with different diffusion coefficients, we have a flow continuity condition as 
\begin{align}\label{BC2}
D_{\rm eff} \frac{\partial c_s(\bar r,t)}{\partial r} =D \frac{\partial c_o(\bar r,t)}{\partial r},
\end{align}
and another boundary condition that is generally modeled as \cite[Ch. 3]{Crank}
\begin{align}\label{BC1}
c_s(\bar r,t)=kc_o(\bar r,t),
\end{align}
where $ \bar r \in \partial \Omega$, $\partial \Omega$ denotes the boundary region of the spheroid, and $c_s$ and $c_o$ denote the concentration function inside and outside the spheroid, respectively. The constant $k$ is a function of porosity of the medium and should be determined experimentally. But, the literature on mathematical analysis of tumor spheroids simply assumes continuity of concentration, i.e., $k=1$, by neglecting the porosity of the medium \cite{Klowss22,Bull20}. However, in the results section, we show using a PBS that for two ideal diffusion environments with diffusion coefficients $D$ and $D_{\rm eff}$, we have $k=\sqrt{\frac{D}{D_{\rm eff}}}$. Thus, for $k\neq 1$, a concentration discontinuity (i.e., jump) occurs at the boundary.
\begin{figure}[!t]
	\centering
	\includegraphics[width=3.7in]{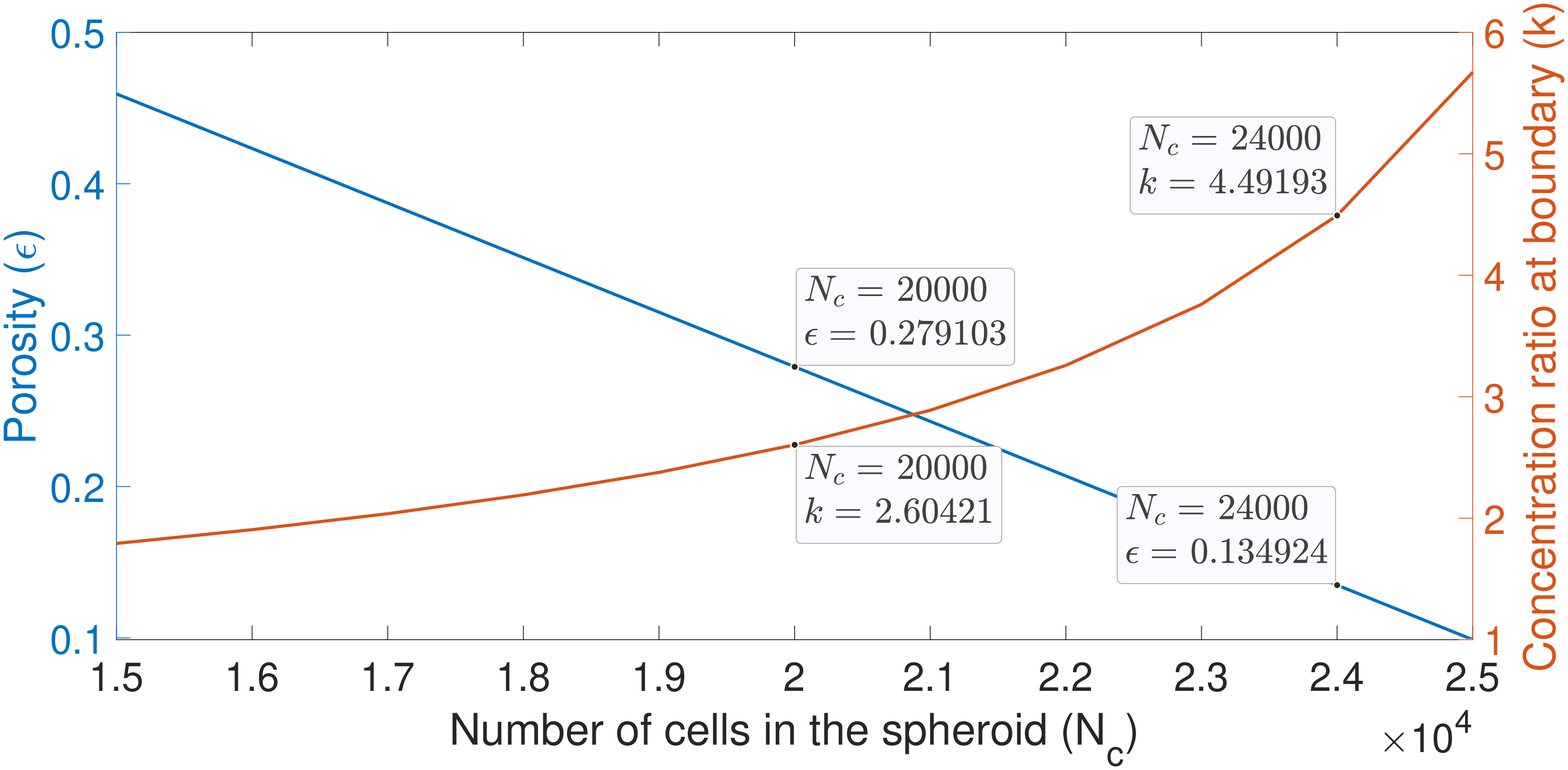}
	\caption{Porosity ($\epsilon$) and concentration ratio at the boundary ($k$) versus the number of cells inside the spheroid ($N_c$).}
	\label{fig1}
\end{figure}

Fig. 1 demonstrates the porosity ($\epsilon$) and boundary concentration ratio ($k$) as a function of the number of the cells, $15000<N_c<25000$ in the spheroid when the spheroid radius and cell volume are assumed to be $R_s=275$ \si{\mu m} and $V_c=3.14 \times 10^{-15}$ \si{m^3}, respectively \cite{Bauer17}. As observed in Fig. 1, the boundary concentration ratio increases exponentially with an increase in $N_c$. For $N_c=24000$, which is the approximate number of cells in HepaRG spheroids reported in \cite{Bauer17}, we have $\epsilon=0.13$ and correspondingly $k=4.49$. This value of k suggests a large concentration discontinuity at the spheroid boundary. 

We assume that the molecules may bind to cell receptors through a chain of reactions leading to product molecules $E$ that are counted by cells, i.e.,
\begin{align}\label{RC1}
\ce{A + B <=>\text{...} -> E},
\end{align}
where $B$ and $E$ may represent the receptor molecules and the final products in the cells, respectively.
%which can be modeled as the set of ordinary differential equations. In Laplace domain, the concentrations of all species are linearly related to the concentration of A molecules.

% a
%\begin{equation}\label{deg1}
%\mathrm{A} \overset{k_{\mathrm d}}{\to} \mathrm{\hat{A}},
%\end{equation}
%where $k_{\mathrm d}$ is the degradation reaction constant in $\mathrm{s}^{-1}$. 
A real mature spheroid is tightly formed by a large number of cells, resulting in a small porosity. Then, a signaling molecule moving within the void space would be very close to the cells' boundaries and frequently exposed to bind to receptors. Thus, we assume that molecules within the void space of the spheroid are homogeneously available for transmembrane reactions.

\subsection{Diffusive MC system with spheroidal receiver}
A point-to-point diffusive MC system is considered in which the spheroid is assumed to be the receiver. The spheroid might model a cancer tumor site that should receive drug molecules; it might model the spheroids in a organ-on-chip system that receive nutrient molecules to survive; alternatively, it could be a transceiver nanomachine for information exchange in a MC system.
  
\begin{figure}[!t]
	\centering
	\includegraphics[width=3in]{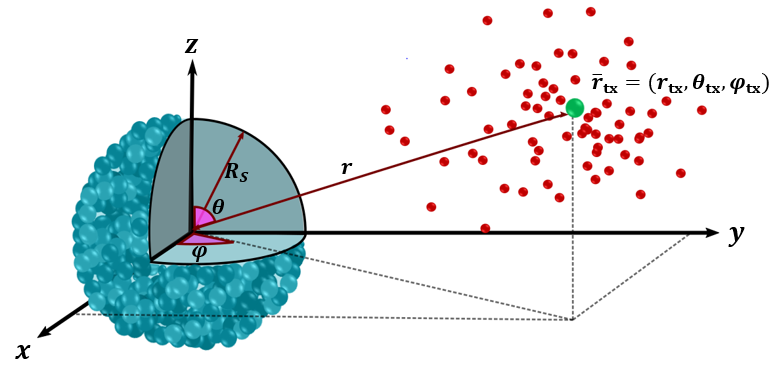}
	\caption{System model schematic where the red, greenish-blue, and green spheres represent the signaling molecule, cell, and point source transmitter, respectively.}
	\label{Fig0}
\end{figure}

 The spheroid is considered in an unbounded fluid medium and the center of the spheroid is chosen as the origin of the spherical coordinate system. A point source transmitter is located at an arbitrary point $\bar{r}_{\rm tx}=(r_{\rm tx},\theta_{\rm tx},\varphi_{\rm tx})$ outside the spheroid. The transmitter uses signaling molecules of type \textit{A}. %The diffusion coefficient of the medium outside the spheroid for information molecule $A$ is denoted by $D$ $\si{m^2.s^{-1}}$.
%Also, a transparent receiver is considered that does not affect the Brownian motion of molecules. The receiver is a sphere with radius $R_{\rm rx}$ whose center is located at $\bar{r}_{\rm rx}=(r_{\rm rx},\theta_{\rm rx},\varphi_{\rm rx})$.
A schematic illustration of the system model is represented in Fig. \ref{Fig0}.
%\begin{figure}
%	\center
%	\includegraphics[width=15 cm,height=9 cm]{EPSfigures/diffusive MC.pdf}
%	\setlength{\abovecaptionskip}{-0.5 cm}
%	\caption{diffusive MC system in biological spherical environment (Only one hemisphere has been illustrated).}
%	\label{Fig0}
%\end{figure}
%Time is divided into time slot durations of $T$ seconds (s). The receiver and transmitter are assumed to be perfectly synchronized \cite{syncArj}. In each time slot, the transmitter releases information molecules into the environment according to the intended symbol. 
The released molecules move randomly in the environment following Brownian motion. Their movements are assumed to be independent of each other. The diffusing molecules, which are exposed to the receptors in the spheroid, may bind and generate molecules $E$ that are counted within the cells in the spheroidal receiver.

The received signal can be defined according to the application. Generally, the cells might respond differently and react to the generated $E$ molecules. A cell's response might be a continuous function of product molecule concentration. Then, the received signal in the spheroid may be defined as the aggregate number of generated $E$ molecules inside the cell. In this case, the concentration profile of $E$ molecules is obtained as a function of signaling $A$ molecules based on \eqref{RC1}. Then, given the uniform density of cells within the spheroid, the received signal is given by 
\begin{align}\label{Re1}
\int_0^{R_s} 4\pi c_E(\bar r,t)V_c \frac{N_c}{V_s} r^2\text{d}r,
\end{align}
where $c_E$ represents the product molecule concentration. Correspondingly, the received signal can be defined as the total generation rate of the product $E$ molecules inside the spheroid, i.e., the time derivative of \eqref{Re1}.
Alternatively, an individual cell might be activated and respond according to a stimulus threshold. For example, if the spheroid is a tumor receiving drug molecules, then each cell might die by receiving a concentration of the drug higher than a threshold. Then, the received signal may be defined as the ratio of killed cells receiving the drug molecules, i.e., 
\begin{align}
\int_0^{R_s} 4\pi I_{c_E(\bar r,t)>\mathcal{T}}V_c \frac{N_c}{V_s} r^2 \text{d}r,
\end{align}
where $I_{(\cdot)}$ is the indicator function and $\mathcal{T}$ is the threshold above which an individual cell is activated.
 %Here, we define the received signal as the whole number of generated molecules $E$ due to binding with the cells receptors in the spheroid. 

\section{Green's Function Boundary Value Problem}
To analyze the presented diffusive MC system, we formulate the corresponding Green's function boundary value problem for diffusion in the environment.
We assume that the point source transmitter, located at an arbitrary point $\bar{r}_{\rm tx}=(r_{\rm tx},\theta_{\rm tx},\varphi_{\rm tx})$ outside the spheroid, has an instantaneous molecule release rate\footnote{Note that the number of molecules does not have a dimension, so we have not represented it in the units.} of $\delta(t-t_0)$ molecule $1/ \rm s$, where $\delta(\cdot)$ is the Dirac delta function. This impulsive point source is represented by the function $S(\bar r,t,{\bar{r}_{\rm tx}},t_0)=\frac{{\delta (r  - {r _{\rm tx}})\delta(\theta-\theta_{\rm tx})\delta(\varphi-\varphi_{\rm tx})\delta(t-t_0)}}{ r^2\sin\theta  }$ $\si{s^{-1}.m^{-3}}$. Given the source $S(\bar r,t,{\bar{r}_{\rm tx}},t_0)$, the molecular diffusion out of the spheroid is described by the partial differential equation (PDE) \cite{Grindrod}
\begin{align}\label{fick}
D{\nabla ^2}c_o(\bar r,t|{{\bar r}_{\rm tx}},{t_0})
 + S(\bar r,t,{\bar{r}_{\rm tx}},t_0) = \frac{{\partial c_o(\bar r,t|{{\bar r}_{\rm tx}},{t_0})}}{{\partial t}},
\end{align}
where $c_o(\bar r,t|{{\bar r}_{\rm tx}},{t_0})$ denotes the molecule concentration at point $\bar r$ outside spheroid at time $t$. In the rest of the paper, we briefly write $c_o$ instead of $c_o(\bar r,t|{{\bar r}_{\rm tx}},{t_0})$.  
In the spherical coordinate system and Fourier domain, \eqref{fick} is re-written as
\small
\begin{align}\label{Eq}
&\frac{D}{{{r^2}}}\frac{\partial }{{\partial r}}\left({r^2}\frac{{\partial C_o}}{{\partial r}}\right) +
\frac{D}{{{r^2}\sin \theta }}\frac{\partial }{{\partial \theta }}\left(\sin \theta \frac{{\partial C_o}}{{\partial \theta }}\right)+ \frac{D}{{{r^2}{{\sin }^2}\theta }}\frac{{{\partial ^2}C_o}}{{\partial {\varphi ^2}}}\\\nonumber
&+\frac{{\delta (r  - {r _{\rm tx}})\delta(\theta-\theta_{\rm tx})\delta(\varphi-\varphi_{\rm tx})\delta(t-t_0)}}{ r^2\sin\theta  } = i\omega C_o, \nonumber
\end{align}
\normalsize
where $C_o$ is the Fourier transform of $c_o$, $i$ is the imaginary unit, and $\omega$ is the angular frequency variable in the Fourier transform\footnote{We use the small $c$ and capital $C$ to represent the concentration function in the time and Fourier domains, respectively.}.
With a first order approximation of the spheroid cell receptor reaction chain (5), the effective diffusion inside the spheroid is modeled as 
\begin{align}\label{fick2}
&D_{\rm eff}{\nabla ^2}C_s(\bar r,\omega|{{\bar r}_{\rm tx}},{t_0})
- \mathcal K (\omega)C(\bar r,\omega|{{\bar r}_{\rm tx}},{t_0})\\
&   = i\omega C_s(\bar r,\omega|{{\bar r}_{\rm tx}},{t_0}),\nonumber
\end{align}
where $\mathcal K(\omega)$ is the net impact on molecules \textit{A} due to the binding process characterized by the reaction chain \eqref{RC1}. 
As mentioned, the boundary conditions at the border of the spheroid are given by \eqref{BC2} and \eqref{BC1}.

%The irreversible ligand-receptor reaction over the sphere boundary given in \eqref{deg2} is characterized by the third type (Robin) boundary condition of \cite{Crank} \footnote{Since the condition is over the inner boundary, i.e.,  $C(\bar r,t|{{\bar r}_{\rm tx}},{t_0})$ is the concentration for $r\leq r_s$, the negative sign on the right side is required.}
%\begin{equation}\label{BD1}
%D\frac{{\partial C(\bar r,t|{{\bar r}_{\rm tx}},{t_0})}}{{\partial r }}\mid_{\bar r=(r_s,\theta,\varphi)}=-k_{\mathrm f} C(r_s,\theta,\varphi,t|{{\bar r}_{\rm tx}},{t_0}).
%\end{equation}
The concentration functions $c_s(\bar r,t|{{\bar r}_{\rm tx}},{t_0})$ and $c_o(\bar r,t|{{\bar r}_{\rm tx}},{t_0})$ that satisfy \eqref{fick} and \eqref{fick2}  subject to the boundary conditions \eqref{BC2} and \eqref{BC1} are called the concentration Green's functions (CGFs) of diffusion inside and outside the spheroid, respectively. We have solved the problem and obtained the series-form expressions \eqref{Eqsd} in the Appendix.

\section{Results}
In this section, we first evaluate the boundary concentration ratio, $k$, with a particle based simulator (PBS). We verify the proposed analysis of the spheroid Green's function using this PBS. Furthermore, we compare the received signal at the spheroidal receiver with a transparent receiver to reveal the signal amplification and dispersion within the spheroid. % and compare with the received signal of a transparent receiver.

The geometric parameters considered in this section are based on the real hepatocyte spheroids used in \cite{Bauer17} where the spheroid's radius is 275 $\mu$m with $N_c\approx 24000$ cells. The volume of one hepatocyte cell is estimated as $3.14 \times 10^{-15}$ \si{m^3}. The reaction \eqref{RC1} inside the spheroid is simply assumed to be an irreversible first order  reaction 
\begin{align}\label{RC2}
\ce{A ->[k_{\mathrm f}] E},
\end{align}
with $k_{\mathrm f}=0.01$ or 0.1 \si{s^{-1}}. Therefore, the generation rate of the $E$ molecules inside the spheroid is obtained as $\frac{{\partial c_E}}{{\partial t}}=k_f c_s$ and correspondingly the term $\mathcal K(\omega)$ in \eqref{fick2} is simply $-k_f$.  

The PBS is implemented in MATLAB (R2021b; The MathWorks, Natick, MA) where the time is divided into time steps of $\Delta t=0.05$ s. The molecules released at the point source transmitter at $\bar r_{\rm tx}=(500\;\si{\mu m},\pi/2,0)$ move independently in the 3-dimensional space. The displacement of a molecule in $\Delta t$ s is modeled as a Gaussian random variable with zero mean and variance $2D\Delta t$ in each dimension of Cartesian coordinates. For the PBS, the spheroid is simply a diffusion environment with effective coefficient $D_{\rm eff}$.
If the calculated displacement vector of a molecule in a time slot passes the spheroid boundary, then the length of the portion of the vector inside the spheroid is updated based on the effective diffusion coefficient $D_{\rm eff}$, i.e., that part of the vector is scaled by the factor $\sqrt{\frac{D_{\rm eff}}{D}}$. 
Considering the reaction \eqref{RC2} inside the spheroid, a molecule may be absorbed from the void space of the spheroid by the cells during a time step $\Delta t$ s with approximate probability $k_{\rm f}\Delta t$ \cite{Elka}. To simulate the impulsive source, $10^6$ molecules are released at the transmitter location. 
To obtain a concentration at a given time and location, we assume a transparent sphere centered at that location with a small radius of 10 \si{\mu m} and count the number of molecules inside the sphere at that time. The concentration would then be the counted number of molecules divided by the volume of the sphere. The concentration value is normalized, i.e., is divided by $10^6$, to obtain the response to the impulsive source. 
\begin{figure}[!t]
	\centering
	\includegraphics[width=3.9in]{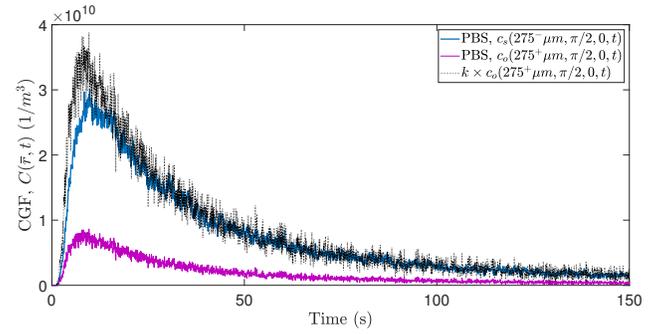}
	\caption{CGF at the inner and outer boundaries versus time obtained from the PBS for $N_c=24000$.}
	\label{Figboundary}
\end{figure}

Fig. \ref{Figboundary} shows the concentration at inner and outer boundary points $c_s(275^- \si{\mu m}, 0, 0)$ and $c_o(275^+ \si{\mu m}, 0, 0)$ of the spheroid given the impulsive source at the transmitter obtained from the PBS when $N_c=24000$. We have also plotted the concentration at the outer boundary scaled by the factor $k$ to verify the boundary condition \eqref{BC1}. The minor mismatch at the peaks is mainly due to the PBS procedure to approximate the concentration at a point. To compute the concentration at the boundary point $(275\;\si{\mu m}, 0, 0)$, we have assumed a transparent sphere of radius $10\; \si{\mu m}$ centered at the boundary point and have counted the molecules in the left and right hemisphere to approximate the concentration of the molecules at inner and outer points. This leads to slightly higher and lower approximations for the concentrations at the outer and inner boundary, respectively.

%\begin{figure}[!t]
%	\centering
%	\includegraphics[width=3.5in]{Figures/Fig3.eps}
%	\caption{Concentrations at the inner and outer boundaries versus time obtained from PBS for $N_c=20000$.}
%	\label{fig1}
%\end{figure}

\begin{figure}[!t]
	\centering
	\includegraphics[width=3.9in]{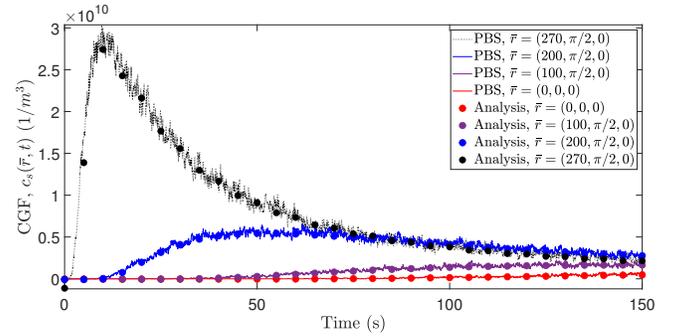}
	\caption{CGF obtained at different points versus time obtained from analysis and PBS for $N_c=24000$.}
	\label{FigCGF}
\end{figure}

Fig. \ref{FigCGF} depicts the concentration at different points inside the spheroidal receiver obtained from the analysis provided in the Appendix and also the PBS when $N_c=24000$. As observed, the PBS fully confirms the analytical results. The figure also indicates that the concentration signal weakens significantly at the points closer to the spheroid center and farther from the point source. In particular, it is observed that the concentration peak at the center is about 18 times smaller than that at the boundary. 

\begin{figure}[!t]
	\centering
	\includegraphics[width=3.9in]{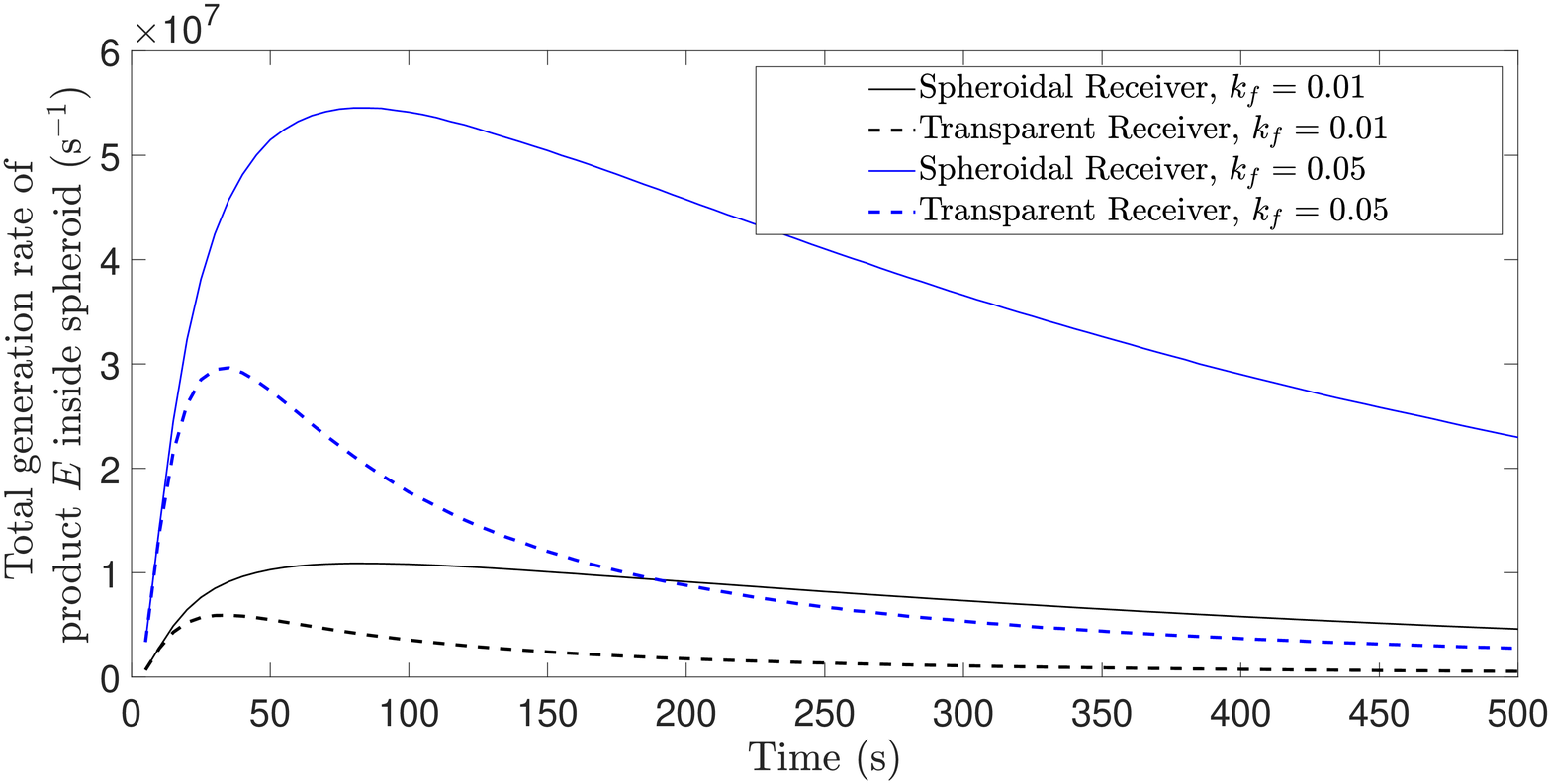}
	\caption{Comparison of the received signal of the spheroidal receiver with the transparent receiver for $N_c=24000$ and $k_f=\{0.01,0.05\}$.}
	\label{FigTran}
\end{figure}

Fig. \ref{FigTran} compares the received signal of the spheroidal and the transparent receivers to show how the spheroid structure affects the diffusion signal. We measure the received signal as the generation rate of the product $E$ molecules all through the spheroid or transparent receiver. Note that the absorbing reaction with the same $k_f$ is also considered within the ``transparent" receiver. As observed, the spheroid structure amplifies the diffusion signal and delays the peak of the received signal, further revealing the dispersion of the diffusion signal inside the spheroid. These two effects of amplification and dispersion lead to a better received signal at the cost of a lower transmission rate that the receiver can perceive. 
\section*{Conclusion}
In this paper, we characterized the received diffusion signal by a spheroidal receiver inspired by spheroids used in organ-on-chip systems. We modeled the spheroid as a porous medium and its boundary condition with free fluid outside. We revealed that the diffusion signal is affected by the spheroid structure in two ways: amplification and dispersion. Both impacts are caused by the effective diffusion coefficient inside the spheroid which is smaller than the diffusion of molecules in the free fluid outside the spheroid. We also provided a series-form expression for the Green's function inside and outside the spheroid and confirmed it using PBS. The proposed spheroid model and analysis not only suggest a new MC receiver, but also support the analysis of organ-on-chip systems where the 3D cell culture spheroids are widely used and the tumors or tumoroids receiving drug molecules in \textit{in vivo} and \textit{in vitro} environments, respectively.    
\section{Acknowledgement}
This work was supported by the Engineering and Physical Sciences Research Council [grant number EP/V030493/1].
%\section*{Acknowledgment}
%\section*{References}
\begin{figure*}
	%\small
	\begin{equation}\label{match}
	\begin{array}{l}
	\sum_{n=0}^{\infty}\sum_{m=0}^{n} H_{mn} \frac{D}{{{r^2}}}\frac{\partial }{{\partial r}}({r^2}\frac{{\partial R_n^o(r,\omega)}}{{\partial r}}) P_{nm}(\cos\theta)\cos(m(\varphi-\varphi_{\rm tx}) +\sum_{n=0}^{\infty}\sum_{m=0}^{n} H_{mn} R_n^o(r,\omega) \frac{D}{{{r^2}\sin \theta }}\frac{\partial }{{\partial \theta }}(\sin \theta \frac{{\partial P_{nm}(\cos\theta)}}{{\partial \theta }}) +\\ \frac{D}{{{r^2}{{\sin }^2}\theta }}\frac{{{\partial ^2}(\cos(m(\varphi-\varphi_{\rm tx})))}}{{\partial {\varphi ^2}}}
	-({i\omega})\sum_{n=0}^{\infty}\sum_{m=0}^{n} H_{mn}R_n^o(r,\omega) P_{nm}(\cos\theta)\cos(m(\varphi-\varphi_{\rm tx}))\\
	= 	-\sum\limits_{n=0}^\infty  \sum\limits_{m = 0}^n 
	L_m \frac{{2n + 1}}{2}\frac{{(n - m)!}}{{(n + m)!}}
	\times{P_{nm}(\cos \theta )P_{nm}(\cos {\theta _{\rm tx}})} {\cos (m(\varphi  - \varphi _{\rm tx}))}\delta(\bar{r}-\bar{r}_{\rm tx}),
	\end{array}
	\end{equation}
	\color{black}
	\hrulefill
\end{figure*}
\begin{figure*}
	\begin{equation}\label{match2}
	\begin{array}{l}
	\sum_{n=0}^{\infty}\sum_{m=0}^{n} H_{mn} \frac{D}{{{r^2}}}\frac{\partial }{{\partial r}}({r^2}\frac{{\partial R_n^s(r,\omega)}}{{\partial r}}) P_{nm}(\cos\theta)\cos(m(\varphi-\varphi_{\rm tx}) +\sum_{n=0}^{\infty}\sum_{m=0}^{n} H_{mn} R_n^s(r,\omega) \frac{D}{{{r^2}\sin \theta }}\frac{\partial }{{\partial \theta }}(\sin \theta \frac{{\partial P_{nm}(\cos\theta)}}{{\partial \theta }}) +\\ \frac{D}{{{r^2}{{\sin }^2}\theta }}\frac{{{\partial ^2}(\cos(m(\varphi-\varphi_{\rm tx})))}}{{\partial {\varphi ^2}}}
	-(\mathcal K(i\omega)+{i\omega})\sum_{n=0}^{\infty}\sum_{m=0}^{n} H_{mn}R_n^s(r,\omega) P_{nm}(\cos\theta)\cos(m(\varphi-\varphi_{\rm tx}))= 	0.
	\end{array}
	\end{equation}
	\color{black}
	\hrulefill
\end{figure*}
\begin{figure*}
	%\small
	\begin{equation}\label{Co}
	\begin{bmatrix}
	j_n(k_1R_s) & -y_n(k_2R_s) & -j'_n(k_2R_s) & 0\\
	D_{\rm eff} k_1 j_n(k_1R_s) & -Dk_2y'_n(k_2R_s) & -Dk_2j'_n(k_2R_s) & 0\\
	0 & y_n(k_2r_{tx}) & j_n(k_2r_{tx}) & -y_n(k_2r_{tx})\\
	0 & r^2_{tx}y'_n(k_2r_{tx}) & r^2_{tx} j'_n(k_2r_{tx}) & -r^2_{tx}y'_n(k_2r_{tx}),
	\end{bmatrix}
	\begin{bmatrix}
	G_n\\
	A_n\\
	B_n\\
	D_n
	\end{bmatrix}
	=
	\begin{bmatrix}
	0\\
	0\\
	0\\
	1
	\end{bmatrix}
	\end{equation}
	%	\hrulefill
\end{figure*}

\appendix 
In this Appendix, we solve the frequency domain PDEs \eqref{Eq} and \eqref{fick2} given the boundary conditions \eqref{BC2} and \eqref{BC1}. The general solution of \eqref{Eq} and \eqref{fick2} is considered as \cite{ZoofaghariWL21} 
\begin{equation}\label{Eqsd}
\begin{aligned}
&C_q(r,\theta ,\varphi ,\omega|\bar{r}_{\rm tx}) 
= \sum\limits_{n = 0}^\infty{\sum\limits_{m = 0}^n {{{H_{mn}R_n^q(r,\omega)}\cos (m(\varphi  - {\varphi _{\rm tx}}))} } } 
\\  &
\times P_{nm}(\cos \theta ),\;\; q\in\{s,o\},
\end{aligned}
\end{equation}
%\begin{figure*}
%	\begin{equation}\label{logender22}
%	\begin{aligned}
%	\frac{D}{{\sin \theta }}\frac{\partial }{{\partial \theta }}\big(\sin \theta \frac{{\partial \mathcal{F}(\theta,\varphi)}}{{\partial \theta }}\big) + \frac{D}{{{{\sin }^2}\theta }}\frac{{{\partial ^2}\mathcal{F}(\theta,\varphi)}}{{\partial {\varphi ^2}}}+{n(n+1)}\mathcal{F}(\theta,\varphi)=0.
%	\end{aligned}
%	\end{equation}
%	\hrulefill
%\end{figure*}  
where  $R_n^q(r,\omega)$  is the unknown radial function of $r$ for the region $q\in\{s,m\}$, $\mathcal{F}(\theta,\varphi)=\cos(m \varphi)P_{nm}(\cos\theta)$ \color{black}
is the associated Fourier-Legendre function of the first kind
with degree $n$ and order $m$ that  satisfies the  partial differential equation (PDE) in
\small 
\begin{equation}\label{logender22}
\begin{aligned}
&\frac{D}{{\sin \theta }}\frac{\partial }{{\partial \theta }}\big(\sin \theta \frac{{\partial \mathcal{F}(\theta,\varphi)}}{{\partial \theta }}\big) + \frac{D}{{{{\sin }^2}\theta }}\frac{{{\partial ^2}\mathcal{F}(\theta,\varphi)}}{{\partial {\varphi ^2}}}\\
&+{n(n+1)}\mathcal{F}(\theta,\varphi)=0,
\end{aligned}
\end{equation}
\normalsize
and $H_{mn}$ denotes the unknown coefficient corresponding to the mode $mn$ of the response. 

Also, the representation of
${\delta (\varphi  - {\varphi _{\rm tx}})} \frac{{\delta (\theta  - {\theta _{\rm tx}})}}{{\sin \theta }}$  based on the aforementioned Fourier-Legendre basis functions is given by  
	\begin{equation}\label{deltaphitheta}
	\begin{aligned}
	&{\delta (\varphi  - {\varphi _{\rm tx}})} \frac{{\delta (\theta  - {\theta _{\rm tx}})}}{{\sin \theta }} =\sum\limits_{n=0}^\infty  \sum\limits_{m = 0}^n L_m \frac{{2n + 1}}{2}\frac{{(n - m)!}}{{(n + m)!}}\\
	%\times
	&{P_n^m(\cos \theta )P_n^m(\cos {\theta _{\mathrm{tx}}})} {\cos \big(m(\varphi  - \varphi _{\rm tx})\big)},
	\end{aligned}
	\end{equation}
%\begin{equation}\label{deltaphitheta}
%\begin{array}{l}
%{\delta (\varphi  - {\varphi _{\rm tx}})} \frac{{\delta (\theta  - {\theta _{tx}})}}{{\sin \theta }} =\sum\limits_{m=0}^\infty  \sum\limits_{n = 0}^\infty 
%L_m \frac{{2n + 1}}{2}\frac{{(n - m)!}}{{(n + m)!}}\\
%\times{P_n^m(\cos \theta )P_n^m(\cos {\theta _{tx}})} {\cos (m(\varphi  - \varphi _{\rm tx}))},
%\end{array}{l}
%\end{equation}
where $L_0=\frac{1}{2\pi}$ and $L_m=\frac{1}{\pi}, m\geq 1$.
Replacing $C_q$ and ${\delta (\varphi  - {\varphi _{\rm tx}})} \frac{{\delta (\theta  - {\theta _{\rm tx}})}}{{\sin \theta }}$  in \eqref{Eq} and \eqref{fick2} by the corresponding series-form representations given by \eqref{Eqsd} and \eqref{deltaphitheta}, respectively, leads to \eqref{match} and \eqref{match2} at the top of this page. Matching the two sides of \eqref{match} yields
\begin{equation}
H_{mn}=L_m \frac{{2n + 1}}{2}\frac{{(n - m)!}}{{(n + m)!}}P_n^m(\cos\theta_{\mathrm{tx}}),
\end{equation} 
and
\begin{equation}\label{RPDE}
\begin{aligned}
r^2\frac{\partial^2 R_n^o(r,\omega)}{\partial r^2}+2r\frac{\partial R_n^o(r,\omega)}{\partial r}&+(k_1^2r^2-n(n+1))R_n^o(r,\omega)\\
&=\delta (\bar r -\bar r_{\mathrm{tx}}),
\end{aligned}
\end{equation}
where  $k_1=\sqrt{\frac{i\omega}{D}}$.
Similarly, \eqref{match2} is reduced to
\begin{equation}\label{RPDE2}
\begin{aligned}
r^2\frac{\partial^2 R_n^s(r,\omega)}{\partial r^2}+2r\frac{\partial R_n^s(r,\omega)}{\partial r}&+(k_2^2r^2-n(n+1))R_n^s(r,\omega)\\
&=0,
\end{aligned}
\end{equation}
where  $k_2=\sqrt{\frac{-{\mathcal K(i\omega)-i\omega}}{D_{\rm eff}}}$. By applying \eqref{Eqsd} in the Fourier forms of  boundary conditions \eqref{BC2} and \eqref{BC1}, we obtain 
\begin{equation}\label{B1}
D_{\rm eff}\frac{\partial R_n^s(r,\omega)}{\partial r }\bigg|_{ r=R_s}=D\frac{\partial R_n^o(r,\omega)}{\partial r }\bigg|_{r =R_s},
\end{equation}
\begin{equation}\label{B2}
R_n^s(r,\omega)\bigg|_{r =R_s}=k\times R_n^o(r,\omega)\bigg|_{r =R_s}.
\end{equation}

To solve \eqref{RPDE}, we remove the source term $\delta (\bar r -\bar r_{\mathrm{tx}})$ at the right-hand side and consider the derivative discontinuity of the CGF at  $r=r_{\mathrm{tx}}$ that leads to the following corresponding boundary condition \cite{ZoofaghariWL21}\footnote{The equivalent boundary condition of \eqref{SD} is derived by integration of two sides of  over $[{r}_{\mathrm{tx}}^{+},{r}_{\mathrm{tx}}^{-}]$.}
\begin{equation}\label{SD}
r^2 \frac{\partial R_n(r,\omega)}{\partial r}\bigg|_{r=r^+_{tx}}
- r^2 \frac{\partial R_n(r,\omega)}{\partial r}\bigg|_{r=r^-_{\mathrm{tx}}}=1.
\end{equation}

The solutions of the homogeneous form of the PDE of \eqref{RPDE} and also \eqref{RPDE2} are then given by
\begin{align}\label{Rr}
&R_n^s(r,\omega) = G_nj_n(kr), \; r<R_s\\
&R_n^o(r,\omega) = \left\{ {\begin{array}{*{20}{c}}
	{\begin{array}{*{20}{c}}
		B_nj_n(kr)+A_n y_n(kr),\;{R_s<r<r_{\rm tx}}
		\end{array}}\\
	{\begin{array}{*{20}{c}}
		D_ny_n(kr),\;{r>r_{\rm tx}}
		\end{array}}
	\end{array}} \right.,
\end{align}
%\begin{equation}\label{Rr}
%R(r)= \{{^{A_nj_n(kr)+B_nn_n(kr)\\\\\\\ r>r_{tx}}_{C_nj_n(kr)+D_nj_n(kr) \\\
%		r<r_{tx}}} 
%\end{equation}
where $j_n(\cdot)$ and $y_n(\cdot)$ are the $n$th order of the first and second types of spherical Bessel function, respectively. %Since $y_n(kr)$ is singular at $r=0$, we set $D_n=0$ for $r<r_{\mathrm{tx}}$. 
By applying the solution \eqref{Rr} to the boundary conditions \eqref{B1}, \eqref{B2}, and \eqref{SD}, and also  the continuity condition of concentration  at $\bar r_{\mathrm{tx}}$, i.e., 
\begin{equation}\label{SC}
R_n^o(r,\omega)\bigg|_{r=r^+_{\mathrm{tx}}}
= R_n^o(r,\omega)\bigg|_{r=r^-_{\mathrm{tx}}},
\end{equation} 
the system of linear equations \eqref{Co} is obtained by which the coefficients $A_n$, $B_n$, $G_n$, and $D_n$ are calculated.
Noteworthy, $j'(\cdot)$ and $y'(\cdot)$ in \eqref{Co} are the derivative functions of $j(\cdot)$ and $y(\cdot)$ in terms of $r$, respectively.  By having the coefficients $A_n,B_n, G_n$, and $D_n$, $R_n^q(r,\omega)$, $q\in\{s,o\}$ in \eqref{Rr} is known and the CGF is computed by taking the inverse Fourier transform of \eqref{Eqsd}.
\balance
%We need to divide the environment into three regions to present the solution: The regions $r<R_s$, $R_s<r<r_{\rm tx}$, and $r>r_{\rm tx}$, where the concentration functions are represented by $C_s$, $C_{m1}$, and $C_{m2}$, respectively. \textbf{Note that the \eqref{Eq} holds for both regions of $R_s<r<r_{\rm tx}$ and $r>r_{\rm tx}$.}
%\vspace{12pt}
\end{document}